\documentclass[a4paper,12pt]{article}
\usepackage[round]{natbib}
\usepackage[english]{babel}
\usepackage{setspace}
\usepackage[utf8x]{inputenc}
\usepackage{graphicx} 
\usepackage[left=2cm,right=2cm,top=2cm,bottom=2cm]{geometry}
\usepackage{textcomp}
\usepackage[T1]{fontenc}
\usepackage{amssymb}
\usepackage{xcolor}
\usepackage{authblk}
\usepackage{ulem}
\usepackage{hyperref}
\usepackage{lineno}
\usepackage{mathabx}
\usepackage{appendix}
\usepackage{float}

\title{
\large
\textbf{Venus boundary layer dynamics: \\ eolian transport and convective vortex}
}

\author[1]{Maxence Lef{\`e}vre}
\affil[1]{Department of Physics (Atmospheric, Oceanic and Planetary Physics), University of Oxford, Oxford, UK}

\date{Accepted in Icarus}

\begin{document}

 \maketitle
\section*{Abstract}

Few spacecraft have studied the dynamics of Venus' deep atmosphere, which is needed to understand the interactions between the surface and atmosphere. Recent global simulations suggest a strong effect of the diurnal cycle of surface winds on the depth of the planetary boundary layer. We propose to use a turbulent-resolving model to characterize the Venus boundary layer and the impact of surface winds for the first time. Simulations were performed in the low plain and high terrain at the Equator and noon and midnight. A strong diurnal cycle is resolved in the high terrain, with a convective layer reaching 7~km above the local surface and vertical wind of 1.3~m/s. The boundary layer depth in the low plain is consistent with the observed wavelength of the dune fields. At noon, the resolved surface wind field for both locations is strong enough to lift dust particles and engender micro-dunes. Convective vortices are resolved for the first time on Venus.

\section{Introduction}

The interaction between the surface and the atmosphere is a major aspect of exchanges of heat and angular momentum, impacting the thermal and wind shear profiles and, therefore, the atmospheric dynamics and the rotation of the solid body itself. On Venus, however, the first 10 km above the surface remain a largely unknown region due to the technological difficulty in probing this region below the cloud layer. 

Only a limited number of probes have been able to collect data on Venus. The VeGa-2 probe has successfully measured the only temperature profile in that region \citep{Link86a}. At the surface, only Venera 9 and 10 directly measured the wind for respectively 49 min and 90~s \citep{Avdu77}, and several other probes like Venera 13 and 14 measured indirectly the wind speed \citep{Ksan83}. The amplitudes of the measured wind speeds are less than 2~m~s$^{-1}$ below 100~m \citep{Lore16}, with a higher probability for values below 0.5~m~s$^{-1}$. The height of the planetary boundary layer (PBL) and the diurnal cycle is not known, nor are the effect of the topography. Dunes have been observed in radar measurements with Magellan \citep{Gree92}, although the lack of knowledge about the spatial and temporal distribution of winds complicates the interpretation of dust transport.

The Institut Pierre Simon Laplace (ISPL) Venus General Circulation Model (GCM) simulations showed the diurnal cycle of the PBL activity is correlated with the diurnal cycle of surface winds \citep{Lebo18}. They observed downward katabatic winds at night and upward anabatic winds during the day along the slopes of high-elevation terrains, resulting in a deeper PBL depth at noon.

\cite{Yama11} performed turbulent-resolving simulations of the Venus PBL with a resolution of 100~m without radiative processes and a variety of near-surface theoretical thermal structures.  This experiment yielded a PBL depth below 2.5~km. \cite{More22} studied the turbulent chemical-species mixing at the surface, where high-density-gradient magnitude regions are formed with larger gradients due to the supercritical conditions.

The PBL of the Earth and Mars have been studied extensively in both observations and modelling studies. On Earth, the turbulent flux is in the energy budget of the convective layer. The opposite is true on Mars. For Venus, this budget is poorly known. On Earth, the PBL dynamic is associated with the presence of the water cycle and latent heat release. The understanding of such dynamics is crucial for the clouds and water cycle and the energy balance of the surface \citep{Garr94}. On Mars and Venus, the latent heat is negligible. The depth of the Martian surface convective layer is larger than on Earth, with also a greater temperature diurnal cycle \citep{Hins08}. The surface slope winds have a strong impact on the thermal structure of the PBL \citep{Spig11}.

In this study, we use Large Eddy Simulation (LES) models developed for the clouds convective regions and to simulate the PBL convective activity at two different locations on the surface. This is done to examine the influence of the topography on the boundary layer convection, and two local times, noon and midnight, to quantify the diurnal cycle. 

For the first time, the radiative processes are taken into account in the study of the Venus PBL with prescribed solar and IR heating rates.  An additional rate representing the effect of the general circulation, heating/cooling due to large-scale wind advection, is also prescribed.

In Section~\ref{Sec:Model1}, the model is described. The spatial and local time variabilities of the PBL are discussed in Section~\ref{Sec:1}. The PBL of Venus is compared to the Earth and Mars in Section~\ref{Sec:3}. The impact of PBL characteristics on eolian transport is discussed in Section~\ref{Sec:4}. Our conclusions are summarized in Section~\ref{Sec:Conc}.

\section{Modelling}

\label{Sec:Model1}

\subsection{Dynamical core}

The LMD LES model is based on the dynamical core of the Advanced Research Weather-Weather Research and Forecast (hereinafter referred to as WRF) terrestrial model \citep{Skam08}. The WRF dynamical core integrates the fully compressible non-hydrostatic Navier-Stokes equations over a specified area of the planet. The conservation of the mass, momentum, and entropy are ensured by an explicitly conservative flux-form formulation of the fundamental equations \citep{Skam08}, based on mass-coupled atmospheric variables (winds and potential temperature). The parametrization of the unresolved small-scale eddies is carried out by a subgrid-scale prognostic Turbulent Kinetic Energy closure by \cite{Dear72}. This methodology has been used for extensively for Earth convection study \citep{Moen07}, and the Martian atmosphere \citep{Spig10}, Venus cloud convective layer \citep{Lefe17,Lefe18}, and terrestrial exoplanet convection \citep{Lefe21}. 

\subsection{Model Physics}

Due to the constant heat capacity of the dynamical core and the architecture of the coupled radiative transfer, as well as computational time, the radiative forcing is handled in the same way that in \cite{Lefe17}. The solar and radiative heating rates are extracted from ISPL Venus GCM \citep{Gara18}, which uses IR transfer \citep{Lebo15} based on \cite{Eyme09}, with the latitudinally-varying cloud model of \citet{Haus14,Haus15}. An additional heating rate is prescribed, representing the large-scale heating from the dynamics. In this study, we focus on the first 10~km and the large-scale heating will come mainly from the anabatic/katabatic slope flows. No surface and sub-surface physics are considered in this study.

\subsection{Simulation settings}
\label{Sec:Model2}

\cite{Lebo18} showed with GCM modelling that the convective depth in the PBL was impacted by the diurnal cycle of the surface wind, and was maximal in the steepest slope of the equatorial topographic features. Therefore, we choose two locations at the surface with two distinct elevations and slope environments at the Equator. Here, the incoming solar flux is maximized in order to study the activity of the PBL where it is supposed to be the most active. One of the locations is in the low plain, with an elevation of -320~m at 0$^{\circ}$ longitude, and the other is in the western part of Ovda Regio with an elevation of 1030~m at 80$^{\circ}$ longitude.
The point in the plain will be hereinafter referred to as \textit{low plain}, and the point in Ovda Regio will be referred to as \textit{high terrain}. The domain of the LES simulations is flat. Due to computational constraints, simulations of an entire Venus day were not possible, and two local times are considered in this study: noon and midnight. The surface is heat flux is set at 90~W~m$^{-2}$ at noon and -1~W~m$^{-2}$ at midnight for the two locations \citep{Lebo18}. This flux is constant in time during the simulations over the entire domain, there is no feedback of the PBL turbulence on the sensible flux. For the two locations, the horizontal resolution and timestep are set at 50~m and 0.4~s. However, the size of the surface area varies depending on local time and location, 30$\times$30~km for the high terrain case at noon and 20$\times$20~km for the rest. The vertical resolution and extent also depend on the location and local time, from 10~km above the local surface with a mean resolution of 90~m for the High terrain case at noon to 5~km above the local surface, with a resolution of 60~m. The different horizontal domains size were chosen to allow several connective cells in each horizontal direction, and were determined by trial and error. The different vertical domains size were chosen to allow several kilometers above the convective layer and were based on the Venus IPSL GCM results \citep{Lebo18}. To avoid the spurious reflection of gravity waves propagating upward on the top of the model, a Rayleigh damping layer is applied over the last 500~m with a damping coefficient of 0.01~s$^{-1}$. The heat capacity is set to a constant value over the whole domain of 1181~J~K$^{-1}$, a reference value from the Venus International Reference Atmosphere \citep{Seif85}.

\begin{figure}[!ht]
 \centering
  \includegraphics[width=17cm]{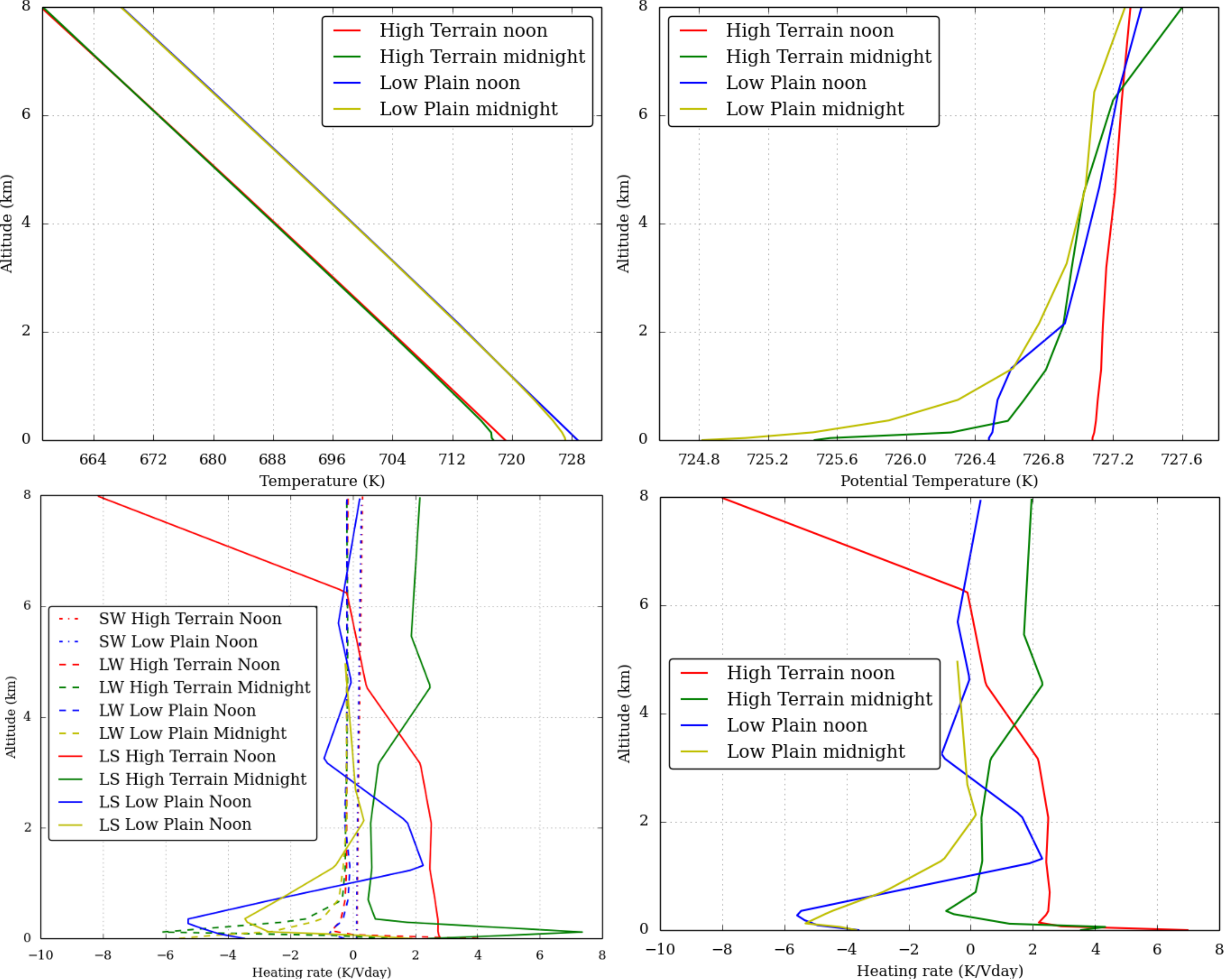}
    \caption{Upper left : initial temperature vertical profiles (K). Upper right : Potential temperature vertical profiles (K). Bottom left: IR (LW), solar (SW) and large-scale (LS) heating rates vertical profiles (K/Venus day). Bottom right : total heating rate vertical profiles (K/Venus day). }
  \label{122}
\end{figure}

Fig~\ref{122} shows the initial profiles of the temperature, potential temperature, and the different heating rates. The temperature is colder for the high terrain cases, and the diurnal cycle of the temperature is below 3~K for the two location cases. At noon, a neutral layer corresponding to the convective layer is visible below 2~km above the local surface for the low plain case, and below 8~km above the local surface for the high terrain case. At midnight, there is no visible neutral layer, meaning that the convective activity is weak. Regarding the heating rates, the short wave heating is slightly greater for the low plain case, but the thermal cooling is slightly stronger at midnight. However, there is a strong difference for the large-scale heating whereas for the high terrain case it is positive up to 6~km above the local surface, with stronger values at noon. While for the low plain case, the large-scale heating is negative in the first 1.5~km at noon and 2~km at midnight, and then alternating between positive and negative values above. This variability of the large-scale heating reflects the effect of the topography and the diurnal cycle of the surface wind. This variability reflects on the total heating rate, alternating between negative and positive values that will enforce the convective depth. The model is initialized with thermal profiles, winds and radiative rates that reached GCM equilibrium, using hypotheses from the subgrid-scale parametrization that will impact the equilibrium state of the region. With the very poor knowledge of the first 12~km, it is difficult to assess the realism of the initial state used for the present studies, although the surface temperature and winds are consistent with measurements. The time step of the simulations is set to 0.4~s, and the radiative transfer is called every 100 dynamical steps, ensuring that there is no issue with numerical precision \citep{Rafk20}. The outputs shown in the following sections are obtained after at least 1 Earth day of simulations for the midnight cases, where a steady-state is reached. For the noon cases, after 2 Earth days, there is a small temperature drift due to the set-up of the surface flux and radiative rates, around 10$^{-8}$~K/s near the surface. A constant forcing corresponding to a single time of day is sometimes used to study self-aggregation in Earth’s tropics \citep{Dale15,Wing17},  as well as tidally-locked rocky exoplanets \citep{Zhan17,Serg20,Lefe21}. When no aggregation is present, the model reaches equilibrium in a couple of days (Wing and Emanuel 2014). The Venus atmosphere can be considered dry. In the present study, there is no water vapor and clouds, and therefore no feedback between moisture and radiation/surface temperature. No aggregation of convection is expected. There is no noticeable change in pressure over time in the domain for all the cases considered. The surface wind's amplitude range does not vary in time. The pseudo-equilibrium is realistic enough to provide qualitative insight into the Venus PBL dynamics.

The model configuration is summarized in Table~\ref{T1}.

\begin{table}[!ht]
\center
\begin{tabular}{|l|cccc|}
\hline
Parameter & \multicolumn{4}{ |c| }{Value} \\
\hline
Gravity (m~s$^{-2}$) & \multicolumn{4}{ |c| }{8.87} \\
\hline
Heat Capacity (J~K$^{-1}$) & \multicolumn{4}{ |c| }{1181} \\
\hline
Surface heat flux (W~m$^{-2}$) & \multicolumn{4}{ |c| }{90 (noon), -1 (midnight)} \\
\hline
Horizontal resolution dx (m) &\multicolumn{4}{ |c| }{50} \\
\hline
Time step (s) &\multicolumn{4}{ |c| }{.4} \\
\hline
Cases & Grid & $\Delta$x and $\Delta$y (km) &$\Delta$z (km) & dz (m)  \\
low plain noon & 401$\times$401$\times$71 & 20 & 5.5 & 80 \\
low plain midnight & 401$\times$401$\times$51 & 20 & 3 & 60 \\
high terrain noon & 601$\times$601$\times$121 & 30 & 10 & 70 \\
high terrain midnight & 401$\times$401$\times$101 & 20 & 5.5 & 55 \\ 
\hline
\end{tabular}
\caption{Planetary and atmospheric parameters for the different simulations. $\Delta$x represents the size of the domain over the axis $x$ and dx the resolution over the axis $x$}
\label{T1}
\end{table}

\section{Spatial and temporal variability of the PBL}
\label{Sec:1}

\begin{figure}[!ht]
 \centering
  \includegraphics[width=17cm]{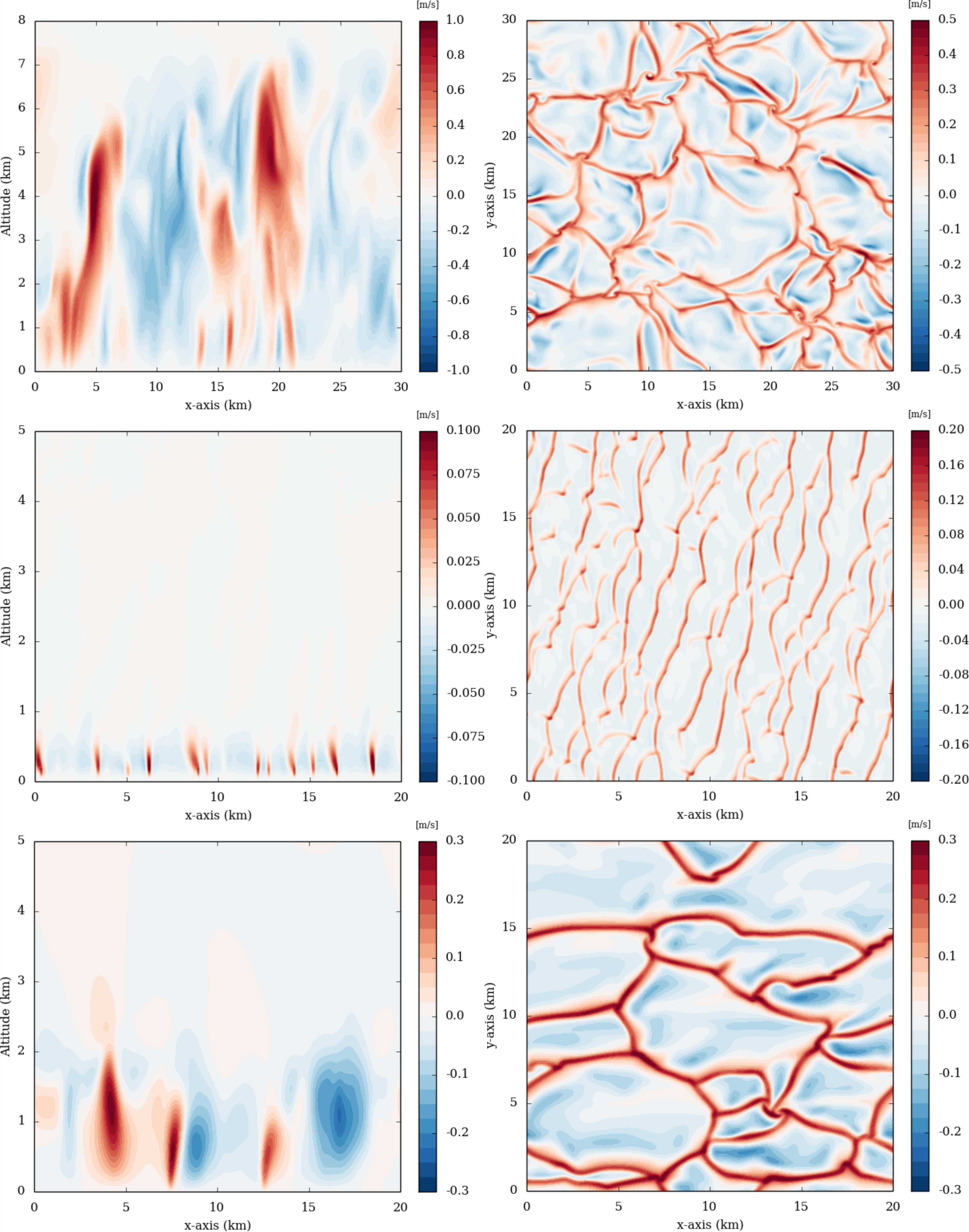}
    \caption{Snapshots of the vertical cross-section (left column row) and horizontal cross (right column) at 100~m above the local surface of the vertical wind (m/s) for the high terrain case at noon (top row) and at midnight (middle row) and the low plain case at noon (bottom column).}
  \label{21}
\end{figure}

Fig~\ref{21} shows snapshots of the vertical and horizontal cross-sections of the vertical wind for the high terrain at noon and midnight, and the low plain at noon. The diurnal cycle for the high terrain location is striking: at noon the elevation of the PBL can reach 7~km above the local surface with vertical wind speed as high as 1.3~m~s$^{-1}$. Whereas at midnight, the convective layer barely reaches 0.5~km with vertical wind speed $<$ 0.2~m~s$^{-1}$.
This diurnal cycle is consistent with GCM simulations \citep{Lebo18}. The difference in depth leads to a difference in convective cell diameter. At noon, the typical cell diameter is around 5~km, with some cells reaching 10~km, and about 2~km at midnight. The cellular features are elongated in the y-direction at midnight. The Richardson number is stronger at midnight than at noon, meaning that the shear becomes relatively strong compared to buoyancy. Such features depend on the vertical shear of the horizontal wind, where there is very little data to constrain  this result. With similar surface flux, the elevation of the PBL depth is different at noon, 2~km above the local surface for the low plain compared to 7~km for the high terrain. This difference is due to the large-scale forcing: in the high terrain this forcing is positive up to 6~km above the local surface (Fig~\ref{122}), bringing energy to the environment and leading to a deeper convective layer. Whereas in the low plain, the large-scale forcing is strongly negative in the first 1.5~km thus stabilizing the atmosphere and leading to a smaller PBL depth. In the high terrain at night, the large-scale forcing will also bring more heat, (although less than at noon) to the environment, but the IR cooling is stabilizing the atmosphere. In the low plain at midnight (not shown here), both the large-scale forcing and IR cooling stabilize the atmosphere, resulting in no convective activity. This PBL spatial variability is also consistent with the GCM simulations \citep{Lebo18}. 

To demonstrate the impact of the large-scale forcing on the convective depth, simulations were carried out without large-scale forcing and wind shear, but with the same thermal profiles. Results are shown in Fig~\ref{A1}. Without the large-scale forcing, the height of the PBL decreases by several kilometers in the high terrain and by several hundred meters in the low plain. At noon, the large-scale circulation will heat the atmosphere. Such an impact on the PBL depth due to the topography  is not observed on Earth, but is on Mars \citep{Spig10}. The convection-resolving model of \cite{Yama11} displays a realistic surface heat flux with a PBL depth of around 2~km, consistent with the low plain case. With higher surface flux, though unrealistic, the convective layer can go as high as 6~km.

\label{A1}
\begin{figure}[!ht]
 \centering
  \includegraphics[width=17cm]{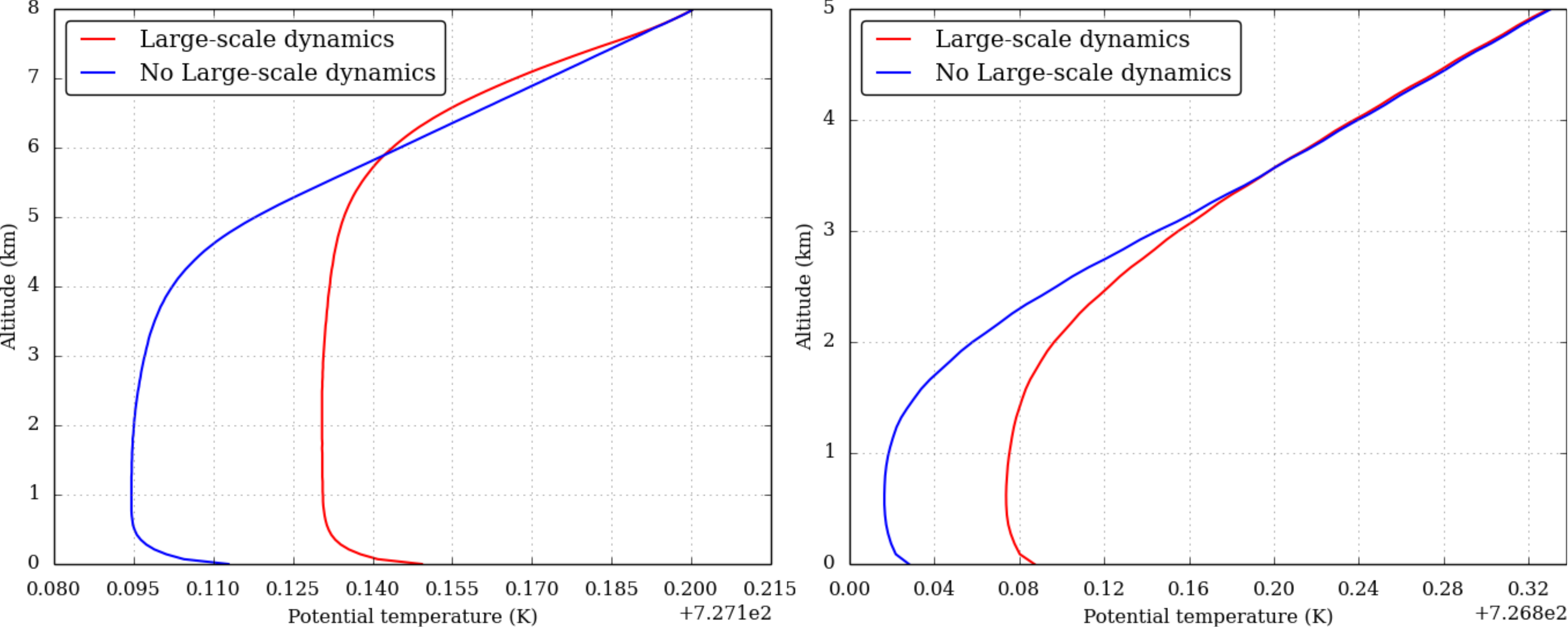}
    \caption{Vertical profile of the domain-average potential temperature for the high terrain case (left) and the low plain case (right) at noon with the large-scale heating rates and wind shear (red lines) and without the large-scale heating rates or wind shear (blue lines). }
  \label{a21}
\end{figure}

In 3D GCMs or 1D models, the turbulence processes are not resolved, therefore a parametrization is needed and a vertical eddy diffusivity coefficient, or $K_{zz}$, is often used to represent convection. In LES models, like the one presented here, the resolution is small enough to resolve the larger eddies, and no vertical eddy diffusivity is therefore needed.

$K_{zz}$ has been estimated in the Venus atmosphere modelling, but is still not well-constrained. The 1D models of \cite{Mats78} and \cite{Taka10} have estimated the vertical eddy diffusion between 10$^2$ and 10$^3$~m$^2$~s$^{-1}$. In the IPSL Venus GCM, the subgrid parametrization exhibits values as high as 1~m$^2$~s$^{-1}$ in the high terrain \citep{Lebo18}. \cite{Yama11} convection-resolving model displays values between 1 and 10$^4$~m$^2$~s$^{-1}$.

Using Prandtl mixing-length theory \citep{Lind71} the vertical eddy diffusion can be estimated from the resolved convective plume and, as expected, it is higher with a stronger convective activity. It reaches values as high as 10$^3$~m$^2$~s$^{-1}$ in the high terrain at noon, consistent with the estimation from \cite{Yama11} with similar PBL depth. However, the values are several orders of magnitude higher than the IPSL Venus GCM. The parametrization of the GCM, based on \cite{Mell82}, does not display realistic values for vertical eddy diffusion compared to the convective-resolving model of \cite{Yama11} and the present study.

\section{Comparison with Earth and Mars}
\label{Sec:3}

\begin{figure}[!ht]
 \centering
  \includegraphics[width=17cm]{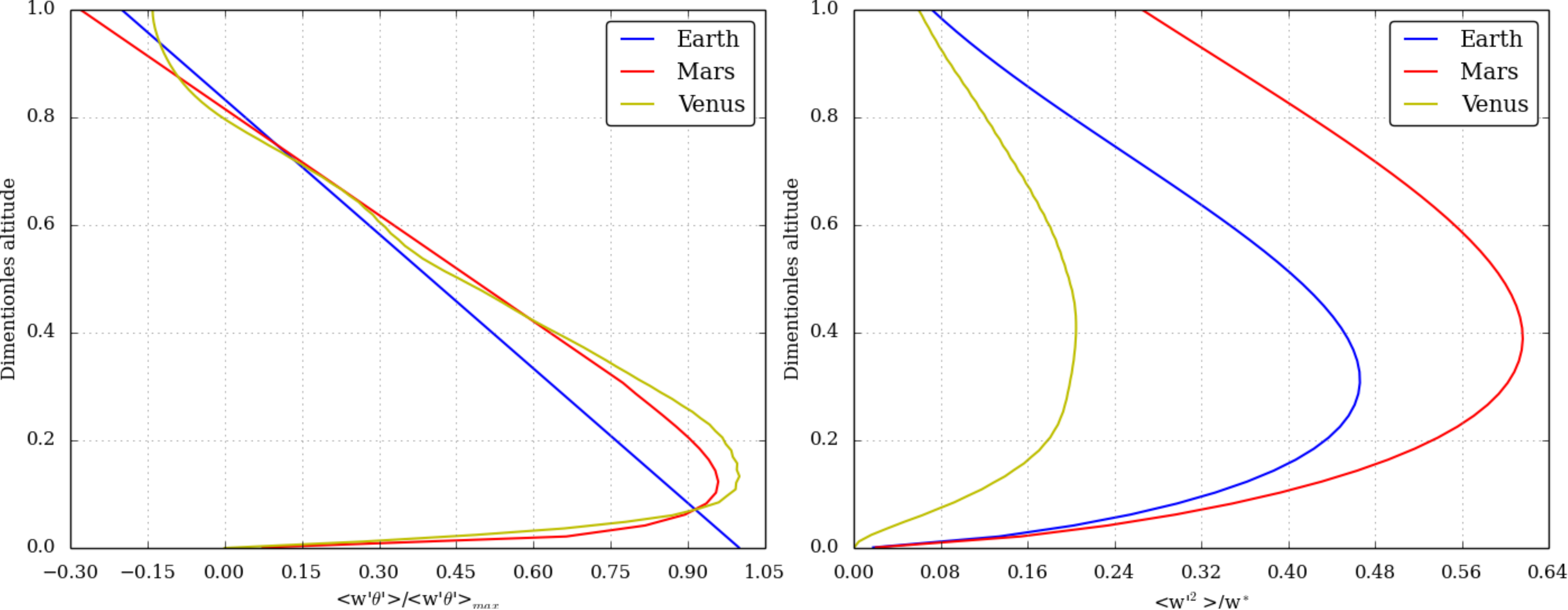}
    \caption{Vertical profiles of the vertical eddy heat flux (left) and vertical velocity variance (right) for Venus, Earth and Mars PBL. Heat flux, variance and height are normalized respectively by the maximum vertical eddy heat flux, the convective velocity scale W$^*$ and the boundary layer depth. For Venus, the high terrain case at noon is presented, Earth data is from \cite{Stul88}, and Mars data is from \cite{Spig10}.}
  \label{41}
\end{figure}

The depth of the convective layer in the low plain at noon is comparable to the Earth's shallow convection depth \citep{Garr94} but with much weaker wind, 0.2~m~s$^{-1}$ for Venus, and several meters per second for Earth \citep{Park18}. The depth of the convective layer in the high terrain is somewhat comparable to the Martin PBL depth, reaching 10~km at maximum \citep{Hins08}. This has a much lower vertical wind amplitude, around 1~m~s$^{-1}$ for Venus, and 10~m~s$^{-1}$ for Mars \citep{Spig10}. The difference in vertical wind is also visible in the convective velocity scale, defined as 

 \begin{equation}
W^{\star}~=~\left[g.z_h \frac{\langle w'\theta' \rangle_{max}}{\langle \theta \rangle}\right]^{1/3}
 \end{equation}
 \label{eq2}
 
with z$_h$ the height of the PBL. For Venus, W$^*$ is around 0.4~m~s$^{-1}$, around 2~m~s$^{-1}$ for Earth \citep{Stul88}, and between 4 and 6.5~m~s$^{-1}$ for Mars \citep{Spig10}. Further comparison between the three atmospheres is shown in Fig~\ref{41} with the vertical eddy heat flux and vertical velocity variance in the PBL in dimensionless defined respectively as 

 \begin{equation}
\frac{\langle w'\theta' \rangle_{max}}{\langle w'\theta' \rangle} \hspace{1cm};\hspace{1cm} \frac{w^{'2}}{W^{\star}}
 \end{equation}
 \label{eq3}

The heat flux on Earth is maximum at the surface, whereas for Venus and Mars fluxes are at the surface, and decrease with height. The turbulence acts to heat the atmosphere. On Mars and Venus, the turbulent flux first increases with height, acting to cool the atmosphere, and then decreases with height, as on Earth, to heat the atmosphere. The low vertical wind amplitude in the PBL shows that with a denser atmosphere, around 50 times higher than the Earth, and 3000 times that of Mars atmospheric density, the heat transport by the atmosphere is more efficient than on other terrestrial planets. For Venus, as on Earth, the energy budget of the convective layer is dominated by the convective flux, with negligible solar and IR heating rates. In the Martian atmosphere, the radiative heating rates are dominant, with surface temperatures that do not depend on altitude, and a pressure effect that is obtained for similar solar heating and IR cooling rates \citep{Spig10}. The empirical similarities in this dimensionless approach between the three planets provide a framework for boundary layer parametrization.

\section{Eolian Transport}
\label{Sec:4}

Only two prominent dune fields have been identified on Venus with the Magellan radar \citep{Gree92}, Algaonice at 25$^{\circ}$S, 340$^{\circ}$E cover some 1300 km~$^2$ with bright dunes of 0.5-5~km in length with a wavelength around 0.5~km, and Fortuna-Meshkenet at 67$^{\circ}$N, 91$^{\circ}$E with transverse dunes of 0.5-10~km long, 0.2-0.5~km wide and spaced by an average of 0.5~km. Several sites with micro-dunes or small-scale wind streaks have also been proposed \citep{Weit94,Bond06} but the resolution of the Magellan radar is not high enough to confirm them.

The wavelength of the observed dune fields is about 500~m. The height of the PBL in Algaonice dune field would be slightly lower than the one at the Equator (Fig~\ref{21}-bottom line) with an averaged depth over a Venus day around 500~m, about the same order of magnitude than the wavelength of the dunes. However, in the Fortuna-Meshkenet the averaged PBL depth would be much lower due to the solar flux more than ten times lower, although the proximity of Ishtar Terra could engender wind turbulence. On Earth, the wavelength of giant dunes is controlled by the depth of the PBL \citep{Andr09}. On Venus, From the two Magellan dune fields observations, the relationship between the wavelength of a giant dune and the depth of the PBL seems to be more complicated, and more observations are needed.

The wind profile is generally approximated by the Prandlt-von K\'arm\'an equation defining the friction velocity u$^{\star}$ as:
 \begin{equation}
u^{\star}~=~k \frac{u_z}{ln(z/z_0)}
\end{equation}
 \label{eq4}
with the Von K\'arm\'an constant, k, equals to 0.4, the surface roughness, z$_0$, set to 1~cm. u$_z$ is the horizontal wind at the altitude z.

Theoretical calculations and laboratory experiments estimated the threshold friction velocity for which the dust is lifted in Venus surface condition to be minimum around 2.5~10$^{-2}$~m~s$^{-1}$ \citep{Iver76}, about 10 times lower than for Earth, and 100 times lower than for Mars due to the very dense atmosphere, depending on dust radius. Such a value depends also on the density of the dust, data that is not well-known yet. Laboratory experiments tested the dust transport in Venus surface conditions and showed that for surface velocities between 0.63 and 1.5~m~s$^{-1}$ there was the formation of micro-dunes, with a wavelength between 8 and 27~cm \citep{Gree84}. Below 0.63~m~s$^{-1}$, the amplitude of the wind is too small to transport dust, and above 1.5~m~s$^{-1}$ no dunes are formed because dust grains can be transported across the lee of a wave onto another and therefore blurring the separation between dunes. 

\begin{figure}[!ht]
 \centering
  \includegraphics[width=17cm]{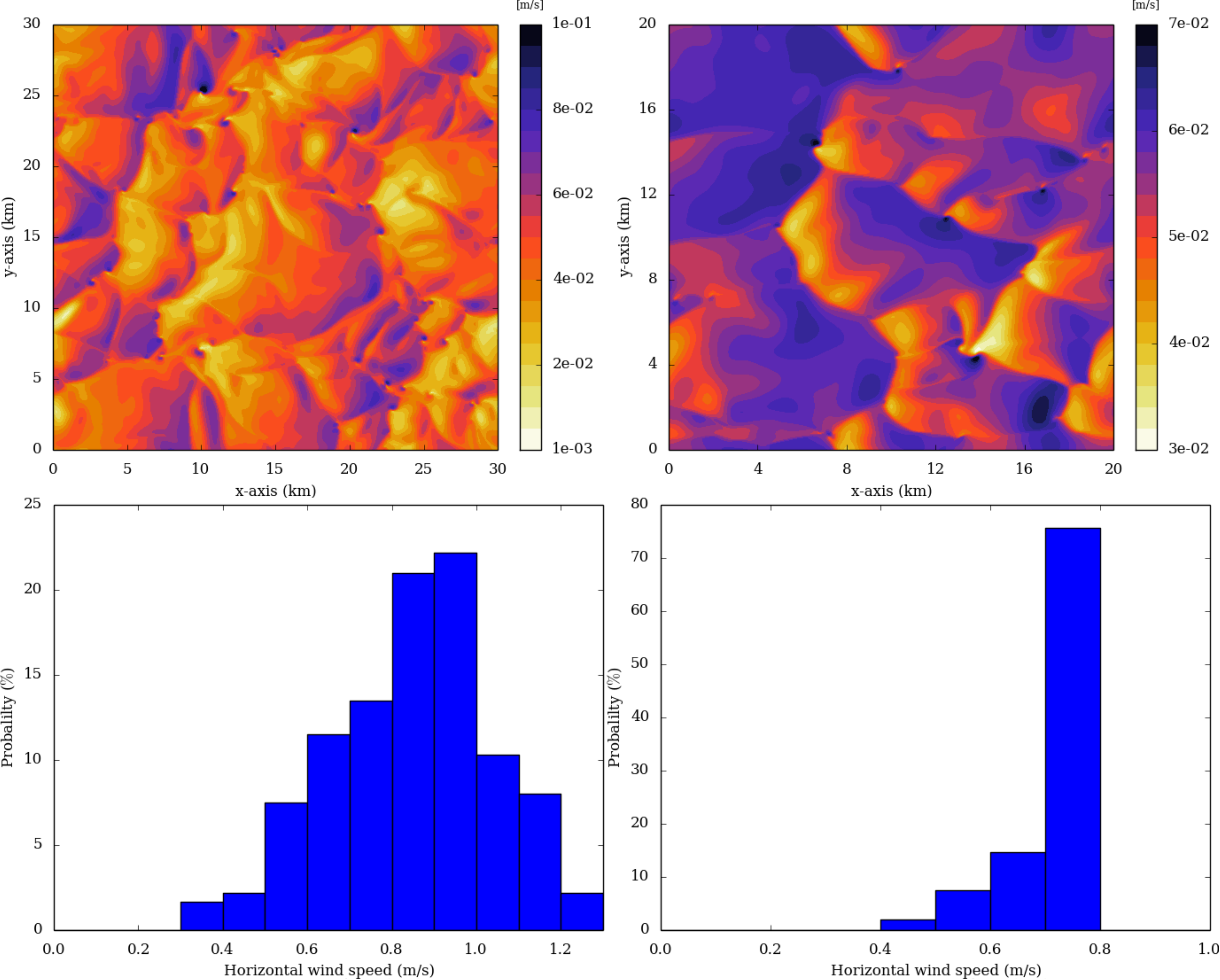}
    \caption{Snapshots of the friction velocity u$^\star$ (top row) and horizontal wind amplitude distribution (bottom row) in the high terrain (left column) and low plain (right column) at 10~m above the local surface, both at noon.}
  \label{51}
\end{figure}

For both the high terrain and low plain, both at noon, the friction velocity of the vast majority of the spatial area is above the saltation threshold, meaning that the turbulence is able to transport dust particles, as shown in Fig~\ref{51}. On the other hand, at night for both locations, the friction velocity is below the saltation threshold over the entire area, and no dust is transported. At night, the horizontal wind amplitude distribution at 10~m above the local surface for both locations is fully consistent with in-situ measurements \citep{Lore16}, with wind speed $<$ 0.5~m~s$^{-1}$. Whereas at noon, the majority of the wind distribution is $>$ 0.5~m~s$^{-1}$, with values of 1~m~s$^{-1}$ for the high terrain. These values are on the outer edge of the observed values, without taking site-specific factors like slopes or local times into account. With a deeper convective layer in the high terrain, the distribution is broadened compared to the low plain location. In the low plain, the horizontal wind amplitude is almost always in the amplitude range where micro-dunes are forming, whereas in the high terrain a significant part of the horizontal wind is too weak to transport wind to form micro-dunes. The formation of micro-dunes is therefore more probable in the low plain at noon, however the spatial distribution of particle reservoir is not known.

\begin{figure}[!ht]
 \centering
  \includegraphics[width=17cm]{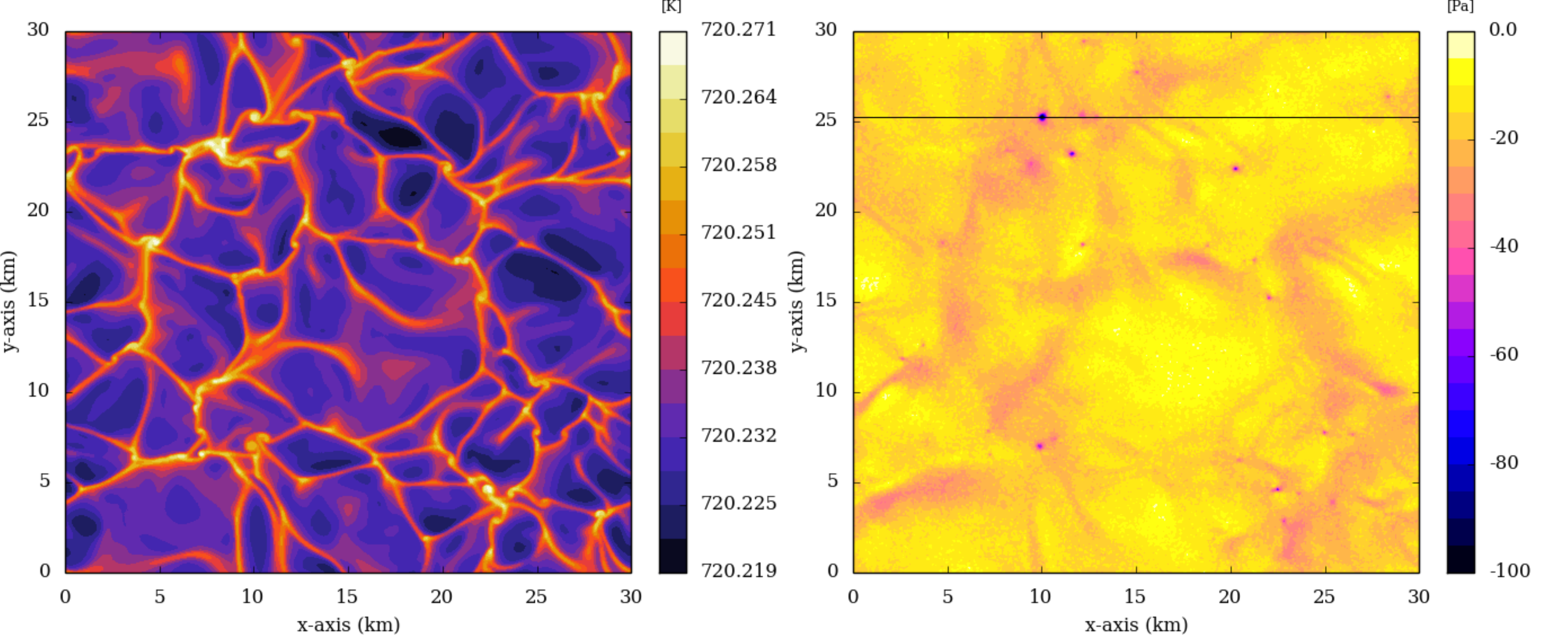}
    \caption{Snapshots of the surface temperature (left), the surface pressure perturbation (right) for the high terrain case at noon. The black line in the right panel represents the cross-section in the following figure.}
  \label{52}
\end{figure}

Fig~\ref{52} shows maps of the surface temperature and surface pressure in the high terrain at noon. The temperature anomaly induced by the convection is very small, around 0.05~K for the strongest convective activity. However, in the pressure field, there is a significant drop, characteristic of convective vortices at the convergence of convective cells \citep{Rafk16}. 

\begin{figure}[!ht]
 \centering
  \includegraphics[width=11.cm]{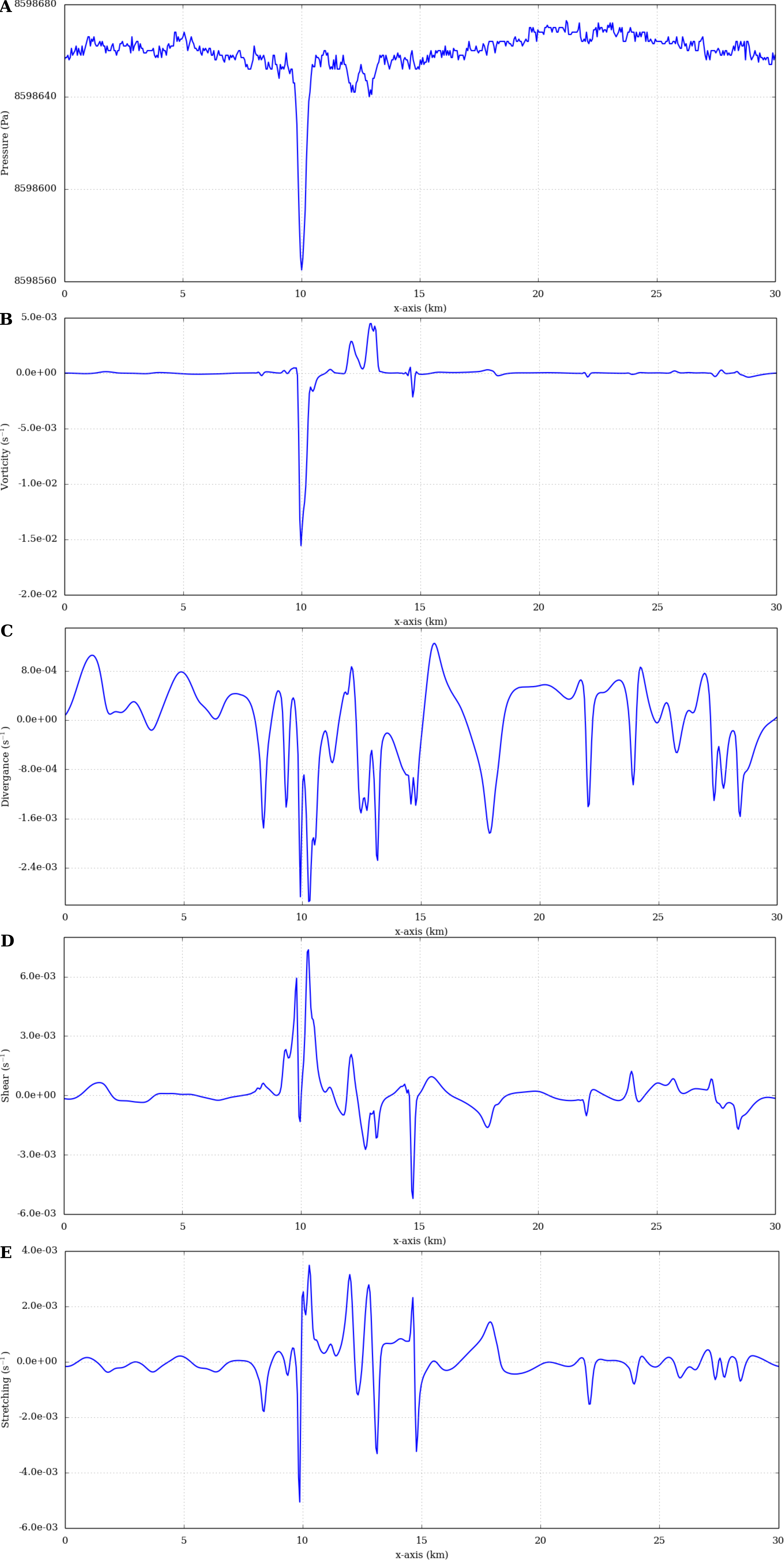}
    \caption{Horizontal cross-sections at y=25.2~km (black line if Fig~\ref{52}) of the surface pressure (A), the vertical component of the rotation vector (B), the horizontal divergence of the horizontal wind (C), the shear (D) and stretching (E) at 10~m above the local surface for the high terrain case at noon.}
  \label{53}
\end{figure}

At these convergences, a pressure deficit is formed from rising plumes, where the updrafts cause a vertical stretching of horizontal wind, leading to a spiralling flow of rising air \citep{Sinc73,Renn98,Toig03,Ito13}. The vorticity inside the vortices, visible with the vertical component of the rotation vector $\zeta$ = ($\partial v$/$\partial x$) - ($\partial u$/$\partial y$) with $u$ and $v$ the two horizontal wind fields. There are vortices with both positive and negative vortices, meaning, respectively, cyclonic and anticyclonic advection. These patterns are also visible in the horizontal divergence $D$ = ($\partial u$/$\partial x$) + ($\partial v$/$\partial y$), with convergence (positive divergence values) for the cyclonic advection and divergence (negative divergence values) for anticyclonic advection. The vortices are also regions where the shear, $\partial u$/$\partial y$ + $\partial v$/$\partial x$, and stretching, $\partial u$/$\partial x$ - $\partial v$/$\partial y$, deformations are strong.  Fig~\ref{53} shows horizontal cross-sections of the surface pressure, vorticity, divergence, shear and stretching at y~=~25.2~km (black line in Fig~\ref{52}) where there is the presence of a strong vortex. The main vortex is visible at x~=~10~km, with the presence of two smaller vortices at x~=~12.2~km and x~=~12.9~km with lower pressure drop. The vortices are strongly visible in the vorticity field, with values up to 2$\times$10$^{-2}$~s$^{-1}$, at least an order of magnitude lower than Earth vortices \citep{Spig16}.The vortices are barely  discernible in the divergence fields, but generate a lot of shear and stretching. There are generations of several strong shear and stretching increase or drops, without noticeable pressure decreases at x~=~14.7~km, 22~km or 23.9~km, for example, due to the presence of thin convective updrafts between convective cells. 

The main vortex extends up to 5~km above the local surface, much higher than on Earth, where it is capped by the height of the PBL (around 2~km). The two smaller vortices only extend up to 1~km above the local surface. The lifetime of such vortices can be several hours. No such structures are resolved in the low plain at noon with a 50~m resolution. The convective activity is either too weak to form vortices, or the resolution of the simulations is too coarse to resolve thin vortices. At night for both locations, no vortices are resolved because the convective activity is too weak to create a vortex. The pressure drops are around 100~Pa for the largest vortices, comparable to the highest pressure drops of dust devils on Earth \citep{Lore14}. However, the diameters are different, almost 1~km on Venus against a few hundreds of meters, maximum, for Earth dust-devils \citep{Lore11}. Convective vortices are also measured in the Martian atmosphere \citep{Bake21}, with pressure drops being inferior to a few pascals \citep{Nish16,Spig21}. 
The 50~m horizontal resolution used for these simulations is sufficient to qualitatively resolve the characteristic structure of convective vortices, but a horizontal resolution one order of magnitude lower would be necessary to resolve a larger part of the spectrum of kinetic energy \citep{Spig16}, as well as smaller vortices. 

\cite{Lore21} used an equation to estimate the maximum vertical wind amplitude in a convective vortex for the different terrestrial planets and Titan. For Venus, this maximum vertical wind amplitude was estimated at 2~m~s$^{-1}$, stronger than the one in the present study around 1.3~m~s$^{-1}$. However, the value of the heat capacity and surface sensitive heat flux chosen in \cite{Lore21} were not representative of the Venus surface at the Equator. With a heat capacity of 1181~J~K$^{-1}$ and a surface sensitive heat flux of 90~W~m$^{-2}$, the estimation with the equation from \cite{Lore21} is 1.35~m~s$^{-1}$, very close to the value obtained in the present model in the high terrain. The location of the vortices is where the friction velocity is the highest and superior to the saltation threshold. Therefore, if there is dust material to be lifted, it will generate dust devils.

\section{Conclusion}
\label{Sec:Conc}

This study presents turbulent-resolving simulations of the Venus PBL, with for the first a realistic thermal profile and with the impact of the large-scale circulation. Simulations were performed at two locations on the Equator, in the low plain and high terrain, and at two local times, noon and midnight. The convective activity is stronger at noon due to the incoming stellar radiation. It is also stronger in the high terrain due to the large-scale heating from the slope winds. The vertical eddy diffusion estimated from turbulent-resolving simulations is not consistent with the GCM, with values several orders of magnitude higher, consistent with previous turbulent-resolving modelling.\\
The PBL of Venus was compared to its equivalent in the atmosphere of the Earth and Mars. On Venus and Mars, the turbulent flux first increases with height, acting to cool the atmosphere, whereas on Earth the turbulence acts to heat the atmosphere. The energy budget of the convective layer is dominated by the convective flux, with negligible solar and IR heating rates for Venus and the Mars. In the Martian atmosphere, the radiative heating rates are dominant.\\
At noon, the turbulent activity is strong enough for the friction velocity to be above the saltation threshold needed for eolian transport. However, there are a lot of unknowns about the composition, density, radius,  and reservoir of such particles that could drastically affect their transport. These simulations were carried out at noon on the Equator, where the convective activity is more intense. More simulations need to be performed at higher latitude and other local times.\\
In the high terrain at noon, the strong activity in the PBL convective vortices are resolved for the first time. Pressure drops as strong as 100~Pa are observed, comparable to the highest pressure drops of dust devils on Earth, but with a larger diameter, almost 1~km on Venus versus a few hundreds km for the Earth. The vertical velocity inside such vortices can exceed 1~m~s$^{-1}$. With the presence of dust, these vortices could lead to dust devils.

\bigbreak
One of the main improvements of the model setup would be to have an interactive surface heat flux and radiative transfer like in the Venus cloud layer \citep{Lefe18}, this would allow feedback between the surface and the atmosphere.

In the IPSL Venus GCM, a nitrogen vertical gradient in the first 7~km from the surface has been tested \citep{Lebo17,Lebo18} and has a substantial impact on the stability of the atmosphere in this region and leads to weaker PBL depth. Such gradient should be tested with a similar setup in future studies. \\

\newpage
\section*{Acknowledgements}
The author would like to thank Aymeric Spiga, S\'ebastien Lebonnois and Kevin Olsen for helpful comments and discussions. The author would like to thank Scot Rafkin and an anonymous reviewer for their help to improve the manuscript. The author acknowledges funding from the European Research Council (ERC) under the European Union’s Horizon 2020 research and innovation program (grant agreement No. 740963/EXOCONDENSE). The author would like to acknowledge the use of the University of Oxford Advanced Research Computing (ARC) facility in carrying out this work. \url{http://dx.doi.org/10.5281/zenodo.22558}.


\begin{thebibliography}{}

\bibitem[{Andreotti} et~al., 2009]{Andr09}
{Andreotti}, B., {Fourri{\`e}re}, A., {Ould-Kaddour}, F., {Murray}, B., and
  {Claudin}, P. (2009).
\newblock {Giant aeolian dune size determined by the average depth of the
  atmospheric boundary layer}.
\newblock {\em Nature}, 457(7233):1120--1123.

\bibitem[{Avduevskii} et~al., 1977]{Avdu77}
{Avduevskii}, V.~S., {Vishnevetskii}, S.~L., {Golov}, I.~A., {Karpeiskii},
  I.~I., {Lavrov}, A.~D., {Likhushin}, V.~I., {Marov}, M.~I., {Melnikov},
  D.~A., {Pomogin}, N.~I., and {Pronina}, N.~N. (1977).
\newblock {Measurement of wind velocity on the surface of Venus during the
  operation of stations Venera 9 and Venera 10}.
\newblock {\em Cosmic Research}, 14(5):710--713.

\bibitem[{Baker} et~al., 2021]{Bake21}
{Baker}, M., {Newman}, C., {Charalambous}, C., {Golombek}, M., {Spiga}, A.,
  {Banfield}, D., {Lemmon}, M., {Banks}, M., {Lorenz}, R., {Garvin}, J.,
  {Grant}, J., {Lewis}, K., {Ansan}, V., {Warner}, N., {Weitz}, C., {Wilson},
  S., and {Rodriguez}, S. (2021).
\newblock {Vortex Dominated Aeolian Activity at InSight's Landing Site, Part 2:
  Local Meteorology, Transport Dynamics, and Model Analysis}.
\newblock {\em Journal of Geophysical Research (Planets)}, 126(4):e06514.

\bibitem[{Bondarenko} et~al., 2006]{Bond06}
{Bondarenko}, N.~V., {Kreslavsky}, M.~A., and {Head}, J.~W. (2006).
\newblock {North-south roughness anisotropy on Venus from the Magellan Radar
  Altimeter: Correlation with geology}.
\newblock {\em Journal of Geophysical Research (Planets)}, 111(E6):E06S12.

\bibitem[{Daleu} et~al., 2015]{Dale15}
{Daleu}, C.~L., {Plant}, R.~S., {Woolnough}, S.~J., {Sessions}, S., {Herman},
  M.~J., {Sobel}, A., {Wang}, S., {Kim}, D., {Cheng}, A., {Bellon}, G.,
  {Peyrille}, P., {Ferry}, F., {Siebesma}, P., and {van Ulft}, L. (2015).
\newblock {Intercomparison of methods of coupling between convection and
  large-scale circulation: 1. Comparison over uniform surface conditions}.
\newblock {\em Journal of Advances in Modeling Earth Systems}, 7(4):1576--1601.

\bibitem[{Deardorff}, 1972]{Dear72}
{Deardorff}, J.~W. (1972).
\newblock {Numerical Investigation of Neutral and Unstable Planetary Boundary
  Layers.}
\newblock {\em Journal of Atmospheric Sciences}, 29:91--115.

\bibitem[{Eymet} et~al., 2009]{Eyme09}
{Eymet}, V., {Fournier}, R., {Dufresne}, J.-L., {Lebonnois}, S., {Hourdin}, F.,
  and {Bullock}, M.~A. (2009).
\newblock {Net exchange parameterization of thermal infrared radiative transfer
  in Venus' atmosphere}.
\newblock {\em J. of Geophys. Res. (Planets)}, 114:E11008.

\bibitem[{Garate-Lopez} and {Lebonnois}, 2018]{Gara18}
{Garate-Lopez}, I. and {Lebonnois}, S. (2018).
\newblock {Latitudinal variation of clouds' structure responsible for Venus'
  cold collar}.
\newblock {\em Icarus}, 314:1--11.

\bibitem[{Garratt}, 1994]{Garr94}
{Garratt}, J.~R. (1994).
\newblock {Review: the atmospheric boundary layer}.
\newblock {\em Earth Science Reviews}, 37(1):89--134.

\bibitem[{Greeley} et~al., 1992]{Gree92}
{Greeley}, R., {Arvidson}, R.~E., {Elachi}, C., {Geringer}, M.~A., {Plaut},
  J.~J., {Saunders}, R.~S., {Schubert}, G., {Stofan}, E.~R., {Thouvenot},
  E.~J.~P., {Wall}, S.~D., and {Weitz}, C.~M. (1992).
\newblock {Aeolian features on Venus - Preliminary Magellan results}.
\newblock {\em Journal of Geophysical Research}, 97:13.

\bibitem[{Greeley} et~al., 1984]{Gree84}
{Greeley}, R., {Marshall}, J.~R., and {Leach}, R.~N. (1984).
\newblock {Microdunes and other aeolian bedforms on Venus: Wind tunnel
  simulations}.
\newblock {\em Icarus}, 60(1):152--160.

\bibitem[{Haus} et~al., 2014]{Haus14}
{Haus}, R., {Kappel}, D., and {Arnold}, G. (2014).
\newblock {Atmospheric thermal structure and cloud features in the southern
  hemisphere of Venus as retrieved from VIRTIS/VEX radiation measurements}.
\newblock {\em Icarus}, 232:232--248.

\bibitem[{Haus} et~al., 2015]{Haus15}
{Haus}, R., {Kappel}, D., and {Arnold}, G. (2015).
\newblock {Radiative heating and cooling in the middle and lower atmosphere of
  Venus and responses to atmospheric and spectroscopic parameter variations}.
\newblock {\em Planetary and Space Science}, 117:262--294.

\bibitem[{Hinson} et~al., 2008]{Hins08}
{Hinson}, D.~P., {P{\"a}tzold}, M., {Tellmann}, S., {H{\"a}usler}, B., and
  {Tyler}, G.~L. (2008).
\newblock {The depth of the convective boundary layer on Mars}.
\newblock {\em Icarus}, 198(1):57--66.

\bibitem[{Ito} et~al., 2013]{Ito13}
{Ito}, J., {Niino}, H., and {Nakanishi}, M. (2013).
\newblock {Formation Mechanism of Dust Devil-Like Vortices in Idealized
  Convective Mixed Layers}.
\newblock {\em Journal of Atmospheric Sciences}, 70(4):1173--1186.

\bibitem[{Iversen} et~al., 1976]{Iver76}
{Iversen}, J.~D., {Greeley}, R., and {Pollack}, J.~B. (1976).
\newblock {Windblown dust on Earth, Mars and Venus.}
\newblock {\em Journal of Atmospheric Sciences}, 33:2425--2429.

\bibitem[{Ksanfomaliti} et~al., 1983]{Ksan83}
{Ksanfomaliti}, L.~V., {Goroshkova}, N.~V., and {Khondyrev}, V.~K. (1983).
\newblock {Wind velocity on the Venus surface from acoustic measurements.}
\newblock {\em Kosmicheskie Issledovaniia}, 21:218--224.

\bibitem[{Lebonnois} et~al., 2015]{Lebo15}
{Lebonnois}, S., {Eymet}, V., {Lee}, C., and {Vatant d'Ollone}, J. (2015).
\newblock {Analysis of the radiative budget of the Venusian atmosphere based on
  infrared Net Exchange Rate formalism}.
\newblock {\em J. of Geophys. Res. (Planets)}, 120:1186--1200.

\bibitem[{Lebonnois} and {Schubert}, 2017]{Lebo17}
{Lebonnois}, S. and {Schubert}, G. (2017).
\newblock {The deep atmosphere of Venus and the possible role of density-driven
  separation of CO$_{2}$ and N$_{2}$}.
\newblock {\em Nature Geoscience}, pages 473--477.

\bibitem[{Lebonnois} et~al., 2018]{Lebo18}
{Lebonnois}, S., {Schubert}, G., {Forget}, F., and {Spiga}, A. (2018).
\newblock {Planetary boundary layer and slope winds on Venus}.
\newblock {\em Icarus}, 314:149--158.

\bibitem[{Lef{\`e}vre} et~al., 2018]{Lefe18}
{Lef{\`e}vre}, M., {Lebonnois}, S., and {Spiga}, A. (2018).
\newblock {Three-Dimensional Turbulence-Resolving Modeling of the Venusian
  Cloud Layer and Induced Gravity Waves: Inclusion of Complete Radiative
  Transfer and Wind Shear}.
\newblock {\em Journal of Geophysical Research (Planets)}, 123:2773--2789.

\bibitem[{Lef{\`e}vre} et~al., 2017]{Lefe17}
{Lef{\`e}vre}, M., {Spiga}, A., and {Lebonnois}, S. (2017).
\newblock {Three-dimensional turbulence-resolving modeling of the Venusian
  cloud layer and induced gravity waves}.
\newblock {\em Journal of Geophysical Research (Planets)}, 122:134--149.

\bibitem[{Lef{\`e}vre} et~al., 2021]{Lefe21}
{Lef{\`e}vre}, M., {Turbet}, M., and {Pierrehumbert}, R. (2021).
\newblock {3D Convection-resolving Model of Temperate, Tidally Locked
  Exoplanets}.
\newblock {\em The Astrophysical Journal}, 913(2):101.

\bibitem[{Lindzen}, 1971]{Lind71}
{Lindzen}, R.~S. (1971).
\newblock {Tides and Gravity Waves in the Upper Atmosphere}.
\newblock In {Fiocco}, G., editor, {\em Mesospheric Models and Related
  Experiments}, volume~25 of {\em Astrophysics and Space Science Library}, page
  122.

\bibitem[{Linkin} et~al., 1986]{Link86a}
{Linkin}, V.~M., {Kerzhanovich}, V.~V., {Lipatov}, A.~N., {Pichkadze}, K.~M.,
  {Shurupov}, A.~A., {Terterashvili}, A.~V., {Ingersoll}, A.~P., {Crisp}, D.,
  {Grossman}, A.~W., {Young}, R.~E., {Seiff}, A., {Ragent}, B., {Blamont},
  J.~E., {Elson}, L.~S., and {Preston}, R.~A. (1986).
\newblock {VEGA balloon dynamics and vertical winds in the Venus middle cloud
  region}.
\newblock {\em Science}, 231:1417--1419.

\bibitem[{Lorenz}, 2011]{Lore11}
{Lorenz}, R. (2011).
\newblock {On the statistical distribution of dust devil diameters}.
\newblock {\em Icarus}, 215(1):381--390.

\bibitem[{Lorenz}, 2016]{Lore16}
{Lorenz}, R.~D. (2016).
\newblock {Surface winds on Venus: Probability distribution from in-situ
  measurements}.
\newblock {\em Icarus}, 264:311--315.

\bibitem[{Lorenz}, 2021]{Lore21}
{Lorenz}, R.~D. (2021).
\newblock {Dust devil winds: Assessing dry convective vortex intensity limits
  at planetary surfaces}.
\newblock {\em Icarus}, 354:114062.

\bibitem[{Lorenz} and {Lanagan}, 2014]{Lore14}
{Lorenz}, R.~D. and {Lanagan}, P.~D. (2014).
\newblock {A Barometric Survey of Dust-Devil Vortices on a Desert Playa}.
\newblock {\em Boundary-Layer Meteorology}, 153(3):555--568.

\bibitem[{Matsuda} and {Matsuno}, 1978]{Mats78}
{Matsuda}, Y. and {Matsuno}, T. (1978).
\newblock {Radiative-convective equilibrium of the Venusian atmosphere}.
\newblock {\em Meteorological Society of Japan}, 56:1--18.

\bibitem[{Mellor} and {Yamada}, 1982]{Mell82}
{Mellor}, G.~L. and {Yamada}, T. (1982).
\newblock {Development of a turbulence closure model for geophysical fluid
  problems}.
\newblock {\em Reviews of Geophysics and Space Physics}, 20:851--875.

\bibitem[Moeng et~al., 2007]{Moen07}
Moeng, C., Dudhia, J., Klemp, J., and Sullivan, P. (2007).
\newblock {Examining Two-Way Grid Nesting for Large Eddy Simulation of the PBL
  Using the WRF Model}.
\newblock {\em Monthly Weather Review}, 135(6):2295--2311.

\bibitem[{Morellina} and {Bellan}, 2022]{More22}
{Morellina}, S. and {Bellan}, J. (2022).
\newblock {Turbulent chemical-species mixing in the Venus lower atmosphere at
  different altitudes: a direct numerical simulation study relevant to
  understanding species spatial distribution}.
\newblock {\em Icarus}, 371:114686.

\bibitem[{Nishizawa} et~al., 2016]{Nish16}
{Nishizawa}, S., {Odaka}, M., {Takahashi}, Y.~O., {Sugiyama}, K.-i.,
  {Nakajima}, K., {Ishiwatari}, M., {Takehiro}, S.-i., {Yashiro}, H., {Sato},
  Y., {Tomita}, H., and {Hayashi}, Y.-Y. (2016).
\newblock {Martian dust devil statistics from high-resolution large-eddy
  simulations}.
\newblock {\em Geophysical Research Letters}, 43(9):4180--4188.

\bibitem[{Park} et~al., 2018]{Park18}
{Park}, S., {Kim}, S.-W., {Park}, M.-S., and {Song}, C.-K. (2018).
\newblock {Measurement of Planetary Boundary Layer Winds with Scanning Doppler
  Lidar}.
\newblock {\em Remote Sensing}, 10(8):1261.

\bibitem[{Rafkin} et~al., 2016]{Rafk16}
{Rafkin}, S., {Jemmett-Smith}, B., {Fenton}, L., {Lorenz}, R., {Takemi}, T.,
  {Ito}, J., and {Tyler}, D. (2016).
\newblock {Dust Devil Formation}.
\newblock {\em Space Science Reviews}, 203(1-4):183--207.

\bibitem[Rafkin and Soto, 2020]{Rafk20}
Rafkin, S.~C. and Soto, A. (2020).
\newblock Air-sea interactions on titan: Lake evaporation, atmospheric
  circulation, and cloud formation.
\newblock {\em Icarus}, page 113903.

\bibitem[{Renn{\'o}} et~al., 1998]{Renn98}
{Renn{\'o}}, N.~O., {Burkett}, M.~L., and {Larkin}, M.~P. (1998).
\newblock {A Simple Thermodynamical Theory for Dust Devils.}
\newblock {\em Journal of Atmospheric Sciences}, 55(21):3244--3252.

\bibitem[{Seiff} et~al., 1985]{Seif85}
{Seiff}, A., {Schofield}, J.~T., {Kliore}, A.~J., {Taylor}, F.~W., {Limaye},
  S.~S., {Revercomb}, H.~E., {Sromovsky}, L.~A., {Kerzhanovich}, V.~V.,
  {Moroz}, V.~I., and {Marov}, M.~Y. (1985).
\newblock {Models of the structure of the atmosphere of Venus from the surface
  to 100 kilometers altitude}.
\newblock {\em Advances in Space Research}, 5:3--58.

\bibitem[{Sergeev} et~al., 2020]{Serg20}
{Sergeev}, D.~E., {Lambert}, F.~H., {Mayne}, N.~J., {Boutle}, I.~A., {Manners},
  J., and {Kohary}, K. (2020).
\newblock {Atmospheric convection plays a key role in the climate of
  tidally-locked terrestrial exoplanets: insights from high-resolution
  simulations}.
\newblock {\em arXiv e-prints}, page arXiv:2004.03007.

\bibitem[{Sinclair}, 1973]{Sinc73}
{Sinclair}, P.~C. (1973).
\newblock {The Lower Structure of Dust Devils.}
\newblock {\em Journal of Atmospheric Sciences}, 30(8):1599--1619.

\bibitem[{Skamarock} and {Klemp}, 2008]{Skam08}
{Skamarock}, W.~C. and {Klemp}, J.~B. (2008).
\newblock {A time-split nonhydrostatic atmospheric model for weather research
  and forecasting applications}.
\newblock {\em Journal of Computational Physics}, 227:3465--3485.

\bibitem[{Spiga} et~al., 2016]{Spig16}
{Spiga}, A., {Barth}, E., {Gu}, Z., {Hoffmann}, F., {Ito}, J., {Jemmett-Smith},
  B., {Klose}, M., {Nishizawa}, S., {Raasch}, S., {Rafkin}, S., {Takemi}, T.,
  {Tyler}, D., and {Wei}, W. (2016).
\newblock {Large-Eddy Simulations of Dust Devils and Convective Vortices}.
\newblock {\em Space Science Reviews}, 203:245--275.

\bibitem[Spiga et~al., 2010]{Spig10}
Spiga, A., Forget, F., Lewis, S.~R., and Hinson, D.~P. (2010).
\newblock Structure and dynamics of the convective boundary layer on mars as
  inferred from large-eddy simulations and remote-sensing measurements.
\newblock {\em Quarterly Journal of the Royal Meteorological Society},
  136:414--428.

\bibitem[{Spiga} et~al., 2011]{Spig11}
{Spiga}, A., {Forget}, F., {Madeleine}, J.-B., {Montabone}, L., {Lewis}, S.~R.,
  and {Millour}, E. (2011).
\newblock {The impact of martian mesoscale winds on surface temperature and on
  the determination of thermal inertia}.
\newblock {\em Icarus}, 212:504--519.

\bibitem[{Spiga} et~al., 2021]{Spig21}
{Spiga}, A., {Murdoch}, N., {Lorenz}, R., {Forget}, F., {Newman}, C.,
  {Rodriguez}, S., {Pla-Garcia}, J., {Moreiras}, D.~V., {Banfield}, D.,
  {Perrin}, C., {Mueller}, N.~T., {Lemmon}, M., {Millour}, E., and {Banerdt},
  W.~B. (2021).
\newblock {A Study of Daytime Convective Vortices and Turbulence in the Martian
  Planetary Boundary Layer Based on Half-a-Year of InSight Atmospheric
  Measurements and Large-Eddy Simulations}.
\newblock {\em Journal of Geophysical Research (Planets)}, 126(1):e06511.

\bibitem[{Stull}, 1988]{Stul88}
{Stull}, R.~B. (1988).
\newblock {\em {An introduction to boundary layer meteorology}}.
\newblock Springer, Dordrecht.

\bibitem[{Takagi} et~al., 2010]{Taka10}
{Takagi}, M., {Suzuki}, K., {Sagawa}, H., {Baron}, P., {Mendrok}, J., {Kasai},
  Y., and {Matsuda}, Y. (2010).
\newblock {Influence of CO$_{2}$ line profiles on radiative and
  radiative-convective equilibrium states of the Venus lower atmosphere}.
\newblock {\em Journal of Geophysical Research (Planets)}, 115(E6):E06014.

\bibitem[{Toigo} et~al., 2003]{Toig03}
{Toigo}, A.~D., {Richardson}, M.~I., {Ewald}, S.~P., and {Gierasch}, P.~J.
  (2003).
\newblock {Numerical simulation of Martian dust devils}.
\newblock {\em Journal of Geophysical Research (Planets)}, 108(E6):5047.

\bibitem[{Weitz} et~al., 1994]{Weit94}
{Weitz}, C.~M., {Plaut}, J.~J., {Greeley}, R., and {Saunders}, R.~S. (1994).
\newblock {Dunes and Microdunes on Venus: Why Were So Few Found in the Magellan
  Data?}
\newblock {\em Icarus}, 112(1):282--295.

\bibitem[{Wing} et~al., 2017]{Wing17}
{Wing}, A.~A., {Emanuel}, K., {Holloway}, C.~E., and {Muller}, C. (2017).
\newblock {Convective Self-Aggregation in Numerical Simulations: A Review}.
\newblock {\em Surveys in Geophysics}, 38(6):1173--1197.

\bibitem[{Wing} and {Emanuel}, 2014]{Wing14}
{Wing}, A.~A. and {Emanuel}, K.~A. (2014).
\newblock {Physical mechanisms controlling self-aggregation of convection in
  idealized numerical modeling simulations}.
\newblock {\em Journal of Advances in Modeling Earth Systems}, 6(1):59--74.

\bibitem[{Yamamoto}, 2011]{Yama11}
{Yamamoto}, M. (2011).
\newblock {Microscale simulations of Venus convective adjustment and mixing
  near the surface: Thermal and material transport processes}.
\newblock {\em Icarus}, 211:993--1006.

\bibitem[{Zhang} et~al., 2017]{Zhan17}
{Zhang}, X., {Tian}, F., {Wang}, Y., {Dudhia}, J., and {Chen}, M. (2017).
\newblock {Surface Variability of Short-wavelength Radiation and Temperature on
  Exoplanets around M Dwarfs}.
\newblock {\em The Astrophysical Journal Letters}, 837:L27.

\end{thebibliography}
\end{document}